\documentclass[aps,prb,reprint]{revtex4-2}
\usepackage[utf8]{inputenc}
\usepackage[english]{babel}
\usepackage{hyperref}
\usepackage{mathtools}
\usepackage{amssymb}
\usepackage[mathscr]{euscript}
\usepackage{bbm}
\usepackage{braket}
\usepackage{xcolor}
\usepackage{graphicx}
\usepackage{tikzit}
\usepackage{tikzscale}

\tikzstyle{round}=[thick, fill=white, draw=black, shape=circle, minimum size=1cm]
\tikzstyle{round conjugate}=[thick, fill=lightgray, draw=black, shape=circle, minimum size=1cm]
\tikzstyle{box}=[thick, fill=white, draw=black, shape=rectangle, rounded corners=.1cm, minimum size=1cm]
\tikzstyle{box conjugate}=[thick, fill=lightgray, draw=black, shape=rectangle, rounded corners=0.1cm, minimum size=1cm]
\tikzstyle{diamond}=[thick, fill=white, draw=black, shape=diamond, minimum size=1.25cm, tikzit shape=rectangle]
\tikzstyle{diamond conjugate}=[thick, fill=lightgray, draw=black, shape=diamond, minimum size=1.25cm, tikzit shape=rectangle]
\tikzstyle{twosite box}=[thick, fill=white, draw=black, shape=rectangle, minimum width=3cm, minimum height=1cm, rounded corners=0.1cm]
\tikzstyle{tall box}=[thick, fill=white, draw=black, shape=rectangle, minimum width=1cm, minimum height=5cm, rounded corners=0.1cm]

\tikzstyle{index}=[-, thick]
\tikzstyle{virtual}=[-, dotted, thick, tikzit draw={rgb,255: red,128; green,128; blue,128}]
\tikzstyle{background}=[-, draw=white, tikzit draw={rgb,255: red,191; green,191; blue,191}, line width=8pt]

\input{tensors.tikzdefs}

\newcommand{\dd}{\text{d}}
\newcommand{\id}{\mathbbm{1}}
\DeclareMathOperator{\real}{Re}

\DeclareMathOperator{\hilb}{\mathscr{H}}
\DeclareMathOperator{\tr}{Tr}
\DeclareMathOperator{\order}{\mathnormal{O}}
\DeclareMathOperator{\Z}{\mathcal{Z}}
\DeclareMathOperator{\dos}{\mathnormal{D}}
\DeclareMathOperator{\lagrangian}{\mathscr{L}}
\DeclareMathOperator{\stiefel}{St}
\DeclareMathOperator{\grassmann}{Gr}
\DeclareMathOperator{\proj}{\mathscr{P}}
\DeclareMathOperator{\retr}{\mathscr{R}}
\DeclareMathOperator{\transp}{\mathscr{T}}
\DeclareMathOperator{\grad}{grad}

\DeclareMathOperator{\J}{\mathscr{J}}

\newcommand{\var}[1]{\braket{\left(\Delta#1\right)^2}}
\newcommand{\pbra}[1]{\langle\!\langle#1|}
\newcommand{\pket}[1]{|#1\rangle\!\rangle}

\newcommand{\includetikz}[1]{\scalebox{0.6}{\input{#1.tikz}}}

\begin{document}

\title{Rényi free energy and variational approximations to thermal states}
\author{Giacomo Giudice}
\affiliation{Max-Planck-Institut f\"{u}r Quantenoptik, Hans-Kopfermann-Stra\ss{}e~1, D-85748 Garching, Germany}
\affiliation{Munich Center for Quantum Science and Technology (MCQST), Schellingstra\ss{}e~4, D-80799 M\"{u}nchen, Germany}
\author{Asl{\i} \c{C}akan}
\affiliation{Max-Planck-Institut f\"{u}r Quantenoptik, Hans-Kopfermann-Stra\ss{}e~1, D-85748 Garching, Germany}
\affiliation{Munich Center for Quantum Science and Technology (MCQST), Schellingstra\ss{}e~4, D-80799 M\"{u}nchen, Germany}
\author{J. Ignacio Cirac}
\affiliation{Max-Planck-Institut f\"{u}r Quantenoptik, Hans-Kopfermann-Stra\ss{}e~1, D-85748 Garching, Germany}
\affiliation{Munich Center for Quantum Science and Technology (MCQST), Schellingstra\ss{}e~4, D-80799 M\"{u}nchen, Germany}
\author{Mari Carmen Ba\~{n}uls}
\affiliation{Max-Planck-Institut f\"{u}r Quantenoptik, Hans-Kopfermann-Stra\ss{}e~1, D-85748 Garching, Germany}
\affiliation{Munich Center for Quantum Science and Technology (MCQST), Schellingstra\ss{}e~4, D-80799 M\"{u}nchen, Germany}
\date{\today}

\begin{abstract}
We propose the construction of thermodynamic ensembles that minimize the Rényi free energy, as an alternative to Gibbs states.
For large systems, the local properties of these Rényi ensembles coincide with those of thermal equilibrium, and they can be used as approximations to thermal states.
We provide algorithms to find tensor network approximations to the 2-Rényi ensemble.
In particular, a matrix-product-state representation can be found by using gradient-based optimization on Riemannian manifolds, or via a non-linear evolution which yields the desired state as a fixed point.
We analyze the performance of the algorithms and the properties of the ensembles on one-dimensional spin chains.
\end{abstract}
\maketitle

\section{Introduction}
\label{sec:intro}
From the point of view of thermodynamics, thermal states describe the equilibrium properties of a system.
Given a Hamiltonian $H$, the Gibbs state
\begin{equation}
  \rho_G = \frac{1}{\Z_G} e^{-\beta H} , \quad \Z_G = \tr e^{-\beta H}
  \label{eq:rho_gibbs}
\end{equation}
describes the state of the system at a given temperature $1/\beta$.
On the other hand, thermal states arise from the principle of maximum entropy~\cite{jaynes1957a,jaynes1957b}: for a given energy, the thermal ensemble is the one that maximizes the von Neumann entropy $S_G(\rho) = - \tr(\rho \log \rho)$.
Equivalently, this can be formulated as the minimization of the free energy
\begin{equation}
  F_{G}(\rho) = \tr(H \rho) - \frac{1}{\beta} S_G(\rho),
  \label{eq:free_energy_gibbs}
\end{equation}
so that
\begin{equation}
  \rho_G (\beta) = \arg \min_{\substack{\rho \succeq 0 \\ \tr \rho = 1}} F_{G}(\rho),
\end{equation}
for some fixed value of $\beta$.
To keep the notation light, we do not explicitly write the dependence of $\rho_G$ on this parameter.
One should keep in mind that the minimum is taken with respect to density operators, i.e.\ positive-semidefinite operators $\rho \succeq 0$ with a chosen normalization, typically $\tr(\rho) = 1$.
This optimization is not very convenient in practice, since the entropy $S_G$ is often difficult to compute, as it requires information about the entire spectrum of $\rho$.
In a quantum many-body setting, this would require diagonalizing an exponentially large operator, because of the inherent tensor product structure of the Hilbert space.

In the quantum many-body setting, numerical approaches to thermal equilibrium do not try to explicitly solve the optimization above, but resort to different approaches to approximate Eqs.~\eqref{eq:rho_gibbs}.
Monte Carlo methods use sampling to estimate very efficiently the physical properties from~Eqs.\eqref{eq:rho_gibbs}, but they encounter difficulties in scenarios where a sign problem appears, as can happen for fermionic models or frustrated systems.
A different approach is based on tensor networks (TNs), where the total state corresponds to the contraction of low-rank tensors and allows for a local description of the physics.
This is motivated by the fact that thermal states for a local Hamiltonian obey an area law for the mutual information~\cite{wolf2008,kuwahara2021}, and hence there is strong theoretical evidence that a tensor network description should be efficient at approximating thermal states~\cite{hastings2006,hastings2006a,molnar2015,guthjarkovsky2020,kuwahara2021}.

In practice, TNs are extremely successful for studying thermal equilibrium.
In one spatial dimension, matrix product states (MPSs) can be used to construct a representation of the (mixed) Gibbs state~\cite{verstraete2004,zwolak2004,feiguin2005,chen2017,chen2018} or, combined with sampling, to construct minimally-entangled thermal states~\cite{white2009,stoudenmire2010}.
Alternatively, the partition function can be represented as a two-dimensional TN, and its contraction can be approximated using tensor renormalization group approaches, for instance, as originally proposed in Refs.~\cite{nishino1995,bursill1996,wang1997}.
The algorithms can also be generalized for two-dimensional systems~\cite{czarnik2012,czarnik2014,kshetrimayum2019}.

In this paper, we study alternative thermodynamic ensembles that, instead of the von Neumann entropy, maximize the $\alpha$-Rényi entropy~\cite{renyi1961},
\begin{equation}
  S_\alpha(\rho) = \frac{1}{1-\alpha}\log{\tr \rho^\alpha}
  \label{eq:entropy_renyi}
\end{equation}
at a fixed energy.
In the limit $\alpha \to 1$, $S_{\alpha}$ reduces to the von Neumann entropy.
By replacing the von Neumann entropy in Eq.~\eqref{eq:free_energy_gibbs} by a Rényi entropy, we define the Rényi free energy:
\begin{equation}
  F_{_\alpha}(\rho) = \tr(H \rho) - \frac{1}{\beta_\alpha} S_\alpha (\rho).
  \label{eq:free_energy_renyi}
\end{equation}
We would like to stress that, in general, the extremizer $\rho_\alpha$ of this function is not the thermal ensemble.
However, as we will show in Sec.~\ref{sec:theory}, this ensemble nonetheless reproduces all local expectation values in the thermodynamic limit.
The parameter $\beta_\alpha$ is not, in general, related to the conventional inverse temperature $\beta$, but should be treated as a constant for the optimization.

From a TN perspective, the definition in Eq.~\eqref{eq:free_energy_renyi} offers the possibility of directly performing a minimization, since the Rényi entropies in Eq.~\eqref{eq:entropy_renyi} are efficiently computable---at least for small integer values of $\alpha$.
In this paper, we analyze the properties of such ensembles, in particular, how they approximate the thermal properties, and present several variational algorithms which can be used to compute them.

For practical purposes, we will often consider the most convenient case $\alpha = 2$, for which Eq.~\eqref{eq:free_energy_renyi} becomes
\begin{equation}
  \rho_R := \arg\min_{\rho \succeq 0 } F_{R}, \quad
  F_{R}(\rho) =  \tr(H \rho) + \frac{1}{\beta_R}\log \tr\rho^2,
  \label{eq:objectivefun}
\end{equation}
where the subscript $R$ represents $\alpha = 2$.
In other words, optimizing Eq.~\eqref{eq:objectivefun} is equivalent to finding the most mixed state at a chosen energy.
In applied mathematics, the optimization of such a function is known as a non-linear semi-definite programming and can be tackled with interior-point methods.
However, in many-body quantum physics, the dimension of $\rho$ increases exponentially with the system size, making such approaches impractical for large systems.

This paper is organized as follows.
In Sec.~\ref{sec:theory}, we provide an analytical solution to Eqs.~\eqref{eq:objectivefun}, expressed in the eigenbasis of the Hamiltonian.
Since the eigenbasis of a many-body system is not always accessible, we propose an optimization strategy based on uniform MPSs, to approximate the purification of $\rho_R$ directly in the thermodynamic limit.
This non-linear optimization can be accelerated using state-of-the-art techniques~\cite{hauru2021} by restricting it to the Grassmann manifold.
This is discussed in detail in Sec.~\ref{sec:optimization}, and accompanying numerical experiments to benchmark the algorithm are presented.
Moreover, we present an alternative technique, based on a non-linear evolution of the density operator in Sec.~\ref{sec:evolution}, which flows toward the desired ensemble.
To conclude, we discuss possible developments in Sec.~\ref{sec:conclusion}.

\section{Theoretical framework}
\label{sec:theory}

\subsection{Maximal Rényi ensemble}
We now show the analytical form of the extremizer of Eq.~\eqref{eq:free_energy_renyi}, which has been previously derived for classical distributions~\cite{bashkirov2004,brody2007,bunte2014}.
We can use this result in the quantum case, noticing that the state that minimizes Eq.~\eqref{eq:free_energy_renyi} must be diagonal in the energy eigenbasis $\{\ket{E_k}\}$ and thus its eigenvalues are equivalent to a probability distribution.

To find the coefficients $\{p_k\}$ in the density operator $\rho = \sum_k p_k \ket{E_k}\bra{E_k}$, $\rho \succeq 0$ which maximizes the Rényi entropy Eq.~\eqref{eq:entropy_renyi} under the constraints
$\tr\rho = 1$ and $\tr(H \rho) = \bar{E}$, we introduce the Lagrange multipliers $\beta_\alpha$ and $\gamma_\alpha$.
The functional $\lagrangian$ is then
\begin{multline}
  \lagrangian(\rho) = \frac{1}{1-\alpha}\log{\sum_k p_k^\alpha} \\-\gamma_\alpha \left(\sum_k p_k - 1\right) - \beta_\alpha \left(\sum_k E_k p_k - \bar{E}\right).
\end{multline}
At the stationary point, the parameter $\gamma_\alpha$ can be eliminated~\cite{bashkirov2004}, and we obtain the \emph{maximal Rényi ensemble} (MRE)
\begin{equation}
  \rho_\alpha = \frac{1}{\Z_\alpha} \Pi_{E_\perp} \left(1 - \beta_\alpha\frac{\alpha - 1}{\alpha}(H - \bar{E}) \right)^{\frac{1}{\alpha - 1}}\Pi_{E_\perp},
  \label{eq:mre}
\end{equation}
where $\Z_\alpha$ is a normalization factor and $\Pi_{E_\perp}$ is a projector onto the eigenvalues below a cutoff energy $E_\perp := \frac{\alpha}{\beta(\alpha - 1)} + \bar{E}$~\footnote{
  Symmetrically, there is also a solution with a projector onto energies above the cutoff energy: $E > E_\perp$.
  For simplicity, we ignore this solution, as it is the Rényi equivalent of negative temperatures.
}:
\begin{align}
  \Pi_{E_\perp} &= \Theta(E_\perp - H), \nonumber \\
  \Z_\alpha &= \tr\left[{\Pi_{E_\perp} \left(1 - \beta_\alpha\frac{\alpha - 1}{\alpha}(H - \bar{E}) \right)^{\frac{1}{\alpha - 1}}}\right],
\end{align}
where $\Theta(\cdot)$ is the Heaviside function.

\begin{figure}[htbp]
  \includegraphics[width=\columnwidth]{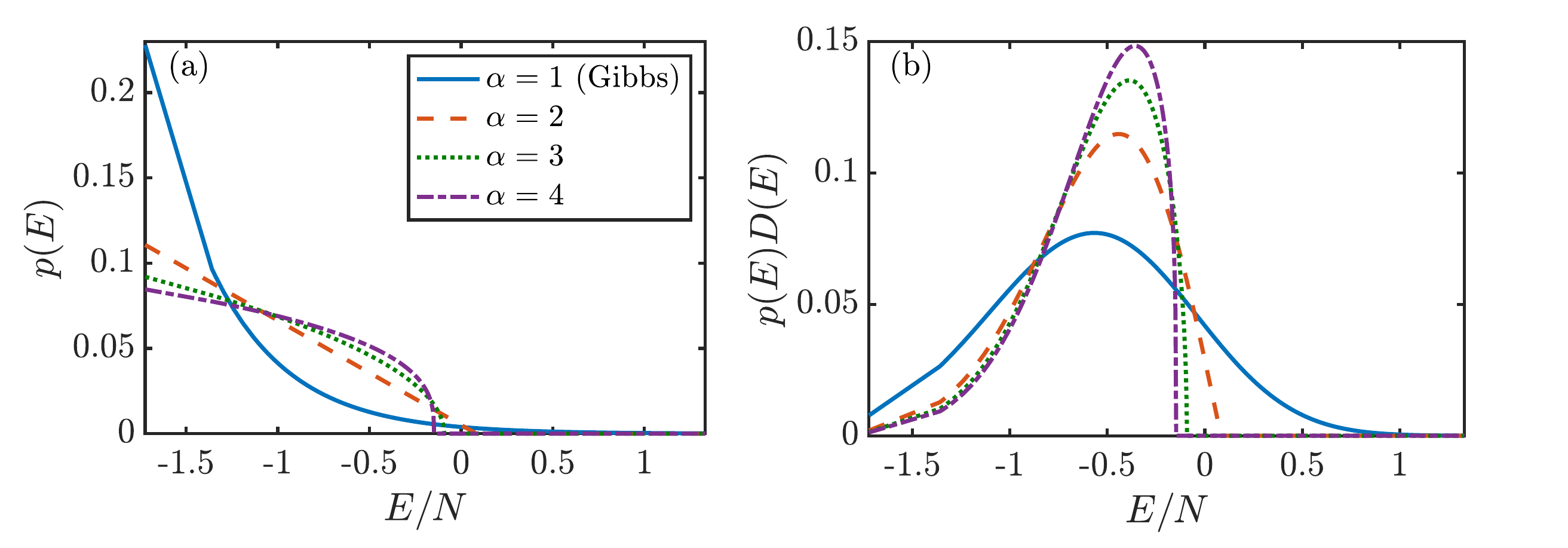}
  \includegraphics[width=\columnwidth]{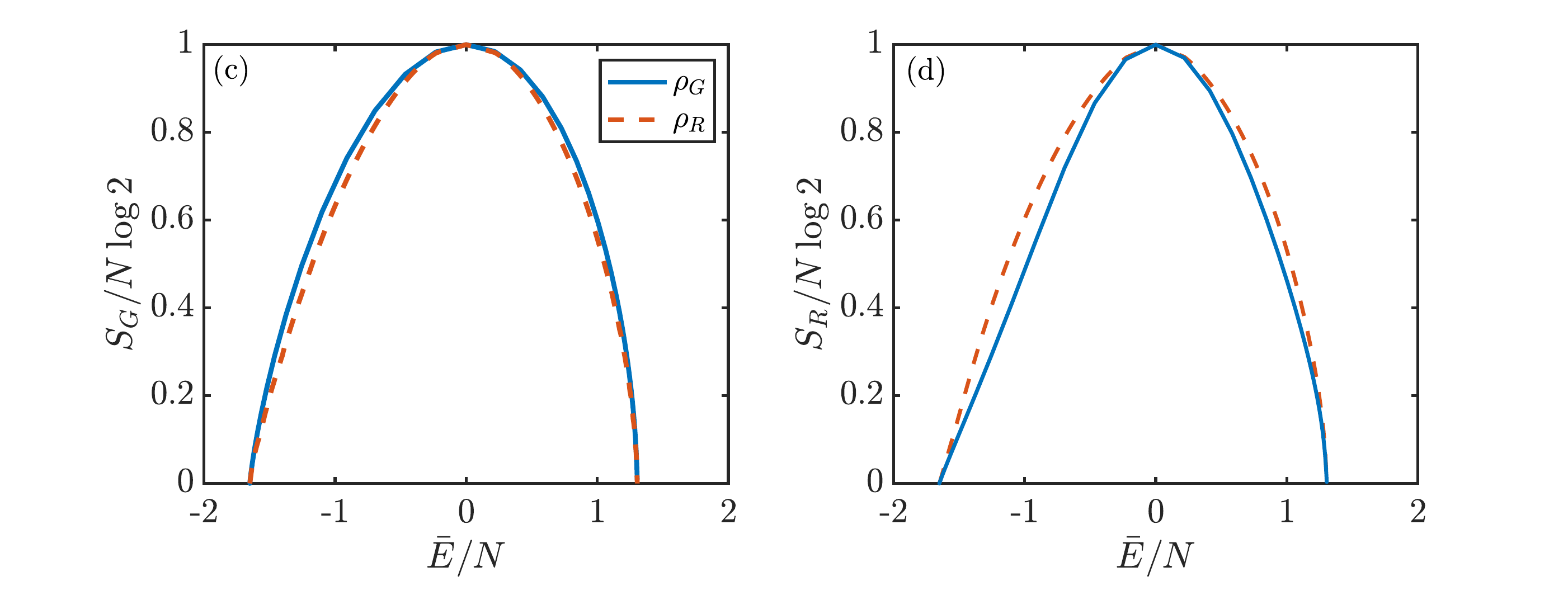}
  \caption{
    (a) Distribution $p(E)$ of the maximal Rényi and Gibbs ensembles for different values of $\alpha$ for the Ising model in Eq.~\eqref{eq:ising}, with longitudinal and transverse fields, respectively, $h_x = 0.5$ and $h_z=-1.05$ and system size $N=10$ (PBC). The mean energy $\bar{E}$ is fixed at $-1/4$ of the width of the spectrum. (b) The same distributions weighted with the corresponding density of states $D(E)$, from the approximation in Ref.~\protect{\cite{atas2014}}.
    Below, the von Neumann (c) and 2-Rényi (d) entropies for the canonical (solid line) and 2-Rényi (dashed line) ensembles are compared at a given mean energy density, for the same system size and Hamiltonian.
    In both cases, the asymptotic behaviors $\lim_{\beta \to 0} S_G = \lim_{\beta_R \to 0} S_R = N \log 2$ and $\lim_{\beta \to \pm\infty} S_G = \lim_{\beta_R \to \pm\infty} S_R = 0$ are recovered.
    The branch with negative (positive) mean energy density corresponds to a $\beta > 0$ ($\beta < 0$), corresponding to a solution with a projector onto energies below (above) the cutoff energy $E_\perp$.
  }
  \label{fig:demo}
\end{figure}

To illustrate the behavior of Eq.~\ref{eq:mre}, we show in Fig.~\ref{fig:demo} some characteristics of the different ensembles in a particular finite case.
In Figs.~\ref{fig:demo}(a) and \ref{fig:demo}(b), we show the distribution of $\rho$ relative to the eigenbasis.
The MRE has a distinctive cutoff energy, beyond which the distribution is zero and therefore fairly different from the case of the canonical ensemble.
However, in a many-body system, we have to consider that the density of states is not uniform but becomes increasingly peaked in the middle of the spectrum.
Then the distributions, weighted by the density of states, become much more similar, as seen in Fig.~\ref{fig:demo}(b).

Another way of visualizing the relation between the canonical and the Rényi ensembles is to compare their entropies for the same mean energy $\bar{E}$.
In Figs.~\ref{fig:demo}(a) and \ref{fig:demo}(b), we explicitly show the comparison of von Neumann and 2-Rényi entropies for the ensembles that maximize each of them over the whole energy range for a small system size. While the behavior is qualitatively similar, both ensembles only coincide in the limiting cases $\bar{E}=0$, when the state is maximally mixed (corresponding to the Gibbs ensemble at infinite temperature $\beta=0$) and $\bar{E}=E_{\min}$ ($E_{\max}$), when the ensemble reduces to the ground (maximally excited) state, corresponding to $\beta \to +\infty$ ($-\infty$).

\begin{figure}[htbp]
  \includegraphics[width=\columnwidth]{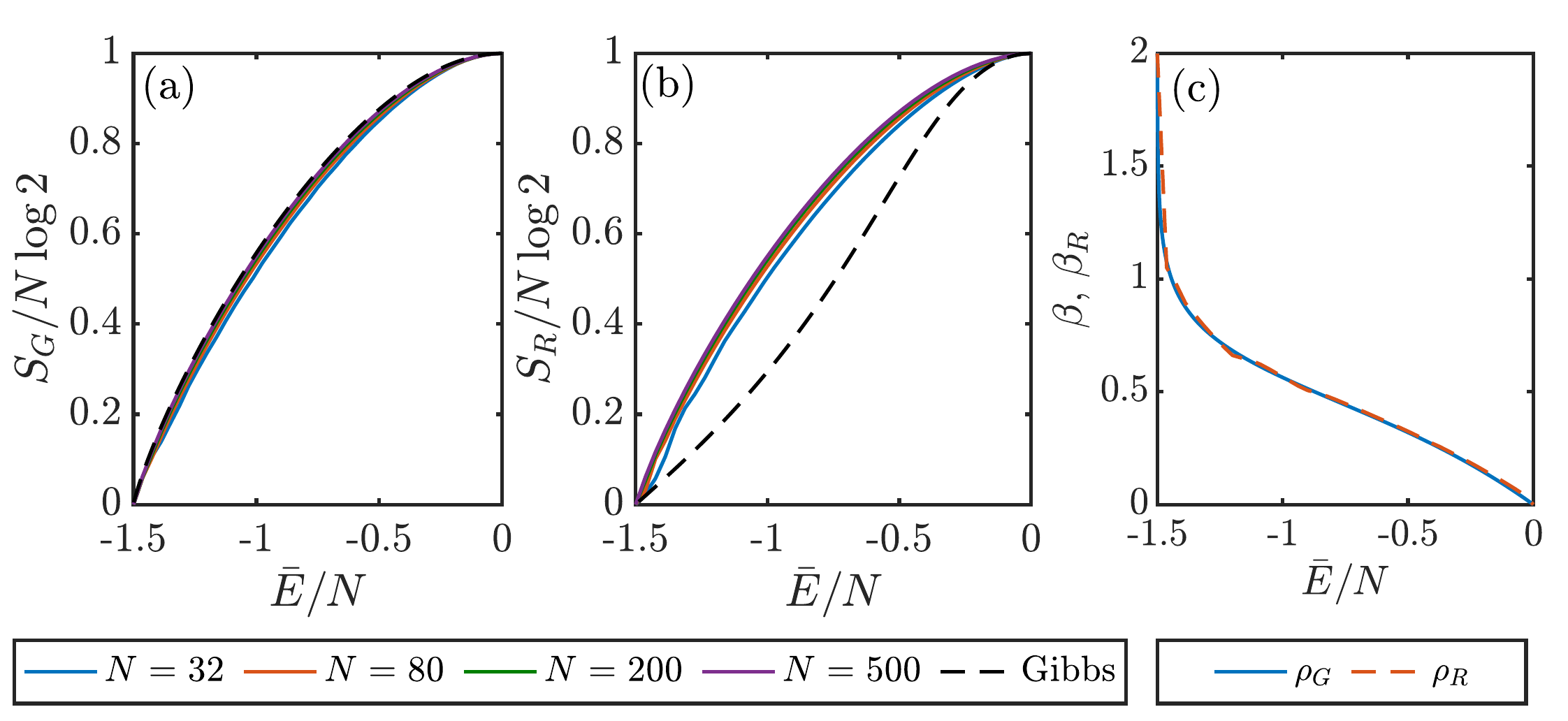}
  \caption{
    (a), (b) Von Neumann and 2-Rényi entropies of $\rho_R$ as a function of mean energy density, for the (classical) Ising model with $h_x = 1/2$, $h_z = 0$.
    Since there is no visible difference in the curves for $\rho_G$, only the largest size ($N=500$) is shown.
    Oscillations at finite size are due to the fact that the eigenvalues correspond to only a number of discrete energies.
    The von Neumann entropy density of the Rényi ensemble approaches that of the Gibbs ensemble, as the system size increases.
    The two ensembles, however, exhibit a difference at intermediate values of the energy density when comparing their 2-Rényi entropy. 
    (c) Comparison of $\beta$ and $\beta_R$ as a function of the mean energy density for the largest size.
    The correspondence is discussed further in Sec.~\ref{sec:descent}.
  }
  \label{fig:classical}
\end{figure}
To study the behavior at large system sizes, we chose to study an exactly solvable case, the results of which are in Fig.~\ref{fig:classical}.
For this Hamiltonian, the density of states becomes Gaussian and the arguments in Appendix~\ref{sec:variance} hold.
While we expect that local observables for both ensembles coincide as the system size increases, the same does not need to hold for non-local quantities, such as the entropies.
It is interesting to notice that the Rényi ensemble has a von Neumann entropy which approaches the Gibbs state, and hence will have a free energy---see Eq.~\eqref{eq:free_energy_gibbs}---which increasingly approaches its maximal value.
However, the same cannot be said for the Rényi free energy introduced in Eq.~\eqref{eq:free_energy_renyi}.

\subsection{Equivalence of local observables}
We now consider a one-dimensional quantum system described by a local Hamiltonian $H$, an operator in the complex Hilbert space $\hilb$.
This total Hilbert space is formed by the tensor product of $N$ local Hilbert spaces:
\begin{equation}
  \hilb = \bigotimes_{n=1}^N \hilb_n.
\end{equation}
The Hamiltonian is restricted to be $\ell$ local, i.e.\ it can then be written in the form
\begin{equation}
  H = \sum_{n=1}^N  h_n,
  \label{eq:hamiltonian}
\end{equation}
where each $h_n$ acts non-trivially only on sites $n,\dots,n+\ell-1$, and has finite operator norm.
Additionally, we will assume that almost all local terms satisfy $\|h_n\|_{\text{op}} > 0$, such that the spectrum of $H$ is extensive.
We mostly consider infinitely large systems, but, when considering finite systems, we specify either open boundary conditions (OBC) or periodic boundary conditions (PBC).

In this setting, it is straightforward to see that the \emph{density of states}
\begin{equation}
  \dos(E) =  \delta (H - E)
  \label{eq:dos}
\end{equation}
has a variance which scales as $\order(\sqrt{N})$.
For specific models, such as strictly one-local Hamiltonians, it can be shown that $D(E)$ becomes Gaussian in the thermodynamic limit.
Under the assumption of a Gaussian density of states, we can then compute the variance of the energy when we take into account the energy distribution of the ensemble.
In the case of the 2-Rényi entropy, it turns out that this can be computed exactly.
As described in Appendix~\ref{sec:variance}, in both cases the variances $\var{H}_G$ and $\var{H}_R$ scale as $\order(N)$.
Hence, if we think about the normalized energy spectrum, both distributions will be increasingly peaked around the same $\bar{E} = \braket{H}$ with a standard deviation $\order(1/\sqrt{N})$ for large $N$.
Hence, the expectation values of local observables become equivalent in the thermodynamic limit.
This derives from the correspondence between microcanonical and canonical ensembles~\cite{landau1991}.
While there exist counterexamples to this correspondence, a sufficient condition for it to hold is that the energy per site converges to a constant~\cite{touchette2003,ellis2004}.
Note that while this argument has been carried out for a Gaussian density of states, we believe that it can be extended to the general case as long as the Hamiltonian is local.

As a final note, we wish to remark that, at least in the case of $\alpha = 2$, we find a correspondence $\beta_R \to \beta$ which holds in the thermodynamic limit.
This holds asymptotically for large $\beta_R$ and the range of validity of this approximation increases with system size.
Hence, the $\beta_R$ for which the Rényi ensemble has the same energy density $\bar{E}$ as a Gibbs ensemble turns out to be the same as the inverse temperature $\beta$.
This can be shown in the case of a Gaussian density of states (see Appendix~\ref{sec:variance}), and is observed numerically in both integrable and non-integrable models (see Sec.~\ref{sec:optimization}).
This is somewhat surprising, since \emph{a priori} there is no connection between the parameters describing the two different ensembles.
From a practical point of view, however, this correspondence is convenient to approximate a thermal ensemble, since we may as well take $\beta_R$ to be the inverse temperature.

\section{Variational algorithms for approximating the Rényi ensemble}
In this section, we introduce two different possibilities to numerically obtain the Rényi ensemble in Eq.~\eqref{eq:mre}.
Although we have a closed form for the exact solution, its use in a many-body setting is impractical because it would require knowledge of the full energy eigenbasis or of the projector in Eq.~\eqref{eq:mre}.
This motivates the formulation of methods compatible with TN techniques.
In Sec.~\ref{sec:optimization}, we explore how uniform MPSs can be used to form a purification which represents the density matrix, and its individual tensors can be optimized directly by using techniques from Riemannian optimization.
In Sec.~\ref{sec:evolution}, instead, we propose a non-linear evolution which has Eq.~\eqref{eq:mre} as a fixed point, so any arbitrary state can be brought to the desired one by simulating this evolution for a sufficient amount of time.

\subsection{Minimization on the MPS manifold}
\label{sec:optimization}
The optimization problem in Eqs.~\eqref{eq:objectivefun} can be restricted to the manifold of states described by some class of tensor networks.
In particular matrix product states (MPS) are arguably the most effective ansatz to represent ground states of local, gapped Hamiltonians in one dimension~\cite{verstraete2006,hastings2006,hastings2007,huang2015}.
We consider a uniform MPS, which written in the conventional diagrammatic notation, is
\begin{multline}
  \ket{\Psi(A)} = \includetikz{mps} \\
  = \sum_{\vec{s}} \tr \left( \dots A^{s_{n-1}} A^{s_{n}} A^{s_{n+1}} \dots \right) \ket{\vec{s}} .
  \label{eq:mps}
\end{multline}
Hence, given a local basis $\{\vec{s}\} = \{s_1, \dots, s_N\}_{s=1,\dots,d}$, each $A$ is a rank-3 tensor with a physical index of dimension $d$ and two virtual indices contracted with the neighboring tensors, each with dimension $D$~\cite{verstraete2008,schollwock2011,orus2014}.
In this section, we focus on uniform MPS for simplicity, but the method can be applied to finite MPSs as well.

A natural generalization of MPS for quantum-mechanical operators is a matrix product operator (MPO)~\cite{verstraete2004,zwolak2004,pirvu2010}, which is composed of rank-4 tensors contracted sequentially,
\begin{equation}
  O = \includetikz{mpo}.
\end{equation}
The issue with this construction is that it is hard to ensure positivity (if the tensors are over the field $\mathbb{C}$ or $\mathbb{R}$), which is a necessary and physical property for objects like density operators.
The problem is that positivity is a global property, which cannot be captured in the local tensors~\cite{delascuevas2013,kliesch2014,delascuevas2016}.
Although an MPO ansatz has been used successfully to approximate the stationary points of dissipative dynamics~\cite{cui2015,mascarenhas2015}, it is problematic for a variational method since there is no way to vary the local tensors without compromising positivity.

An alternative is to introduce a \emph{locally purified state}~\cite{verstraete2004,delascuevas2013,werner2016}, which guarantees the positivity of the operator for any local tensor.
The construction goes as follows.
One considers a pure state, where each site has twice the degrees of freedom, which we call \emph{system} and \emph{ancilla}, so the local tensor is
\begin{equation}
  A = \includetikz{purification-tensor}.
\end{equation}
By tracing out the ancillary degrees of freedom, we obtain a ladder-like TN, which represents the density matrix $\rho = \tr_{\mathrm{anc}} \ket{\Psi}\bra{\Psi}$, or, graphically:
\begin{equation}
  \rho = \includetikz{construction}.
\end{equation}
Shaded boxes represent complex conjugation.
It is simple to see that this TN is positive semidefinite by construction.
The price to pay is that we have introduced a non-linearity in $\rho$ with respect to the local tensors $A^{s}$, so even if the objective function is quadratic in $\rho$, as in Eqs.~\eqref{eq:objectivefun}, it will be quartic in the local tensors.
Hence we cannot use linear algebra to iteratively optimize the local tensors, as in the case of the density matrix renormalization group (DMRG)~\cite{schollwock2011}.
Nonetheless, we can consider the problem in Eq.~\eqref{eq:objectivefun} a non-linear optimization over the tensors of an MPS.

The parametrization of the state in Eq.~\eqref{eq:mps} has an inherent redundancy, since we can perform a gauge transformation on the virtual degrees of freedom of the form $A^s \mapsto X A^s X^{-1}$, for any invertible matrix $X$.
This gauge redundancy of the MPS parametrization allows us to choose the tensors to fulfill the left-gauge condition:
\begin{equation}
  \includetikz{fixedpoint-left}, \quad \includetikz{fixedpoint-right}.
  \label{eq:fixedpoints}
\end{equation}
For the rest of this paper, we will often not draw the ancillary degree of freedom, but implicitly assume it is part of the physical leg of each tensor.
The tensor $\varrho$ is the (positive-semidefinite) right fixed point of the transfer matrix, which encodes the Schmidt values~\cite{schollwock2011}.
Hence, we can view the tensor $A$ as a linear map from the right virtual leg to the left virtual and
physical legs, which is \emph{isometric}.
We will use $W$: $\mathbb{C}^D \mapsto \mathbb{C}^d \times \mathbb{C}^D$ to denote this specific mapping.
Alternatively, we can think of $W$ as a matrix, so we can use the notation $W^\dag W = \id$ to unambiguously specify the isometricity condition.

Hence we can restrict a generic optimization of an MPS to the optimization of tensors over the Stiefel manifold~\cite{absil2008},
\begin{equation}
  \stiefel(n,p) = \left\{W \in \mathbb{C}^{n \times p} | W^\dag W = \id \right\}.
\end{equation}
In reality, since there is a unitary freedom remaining in Eq.~\eqref{eq:fixedpoints}---namely, $A^s \mapsto U^\dag A^s U$---one can restrict the manifold even further to the Grassmann manifold.
The Grassmann manifold should be understood as a quotient manifold, namely all $W$ satisfying the isometricity condition up to a basis rotation, and it is often denoted as $\grassmann(n,p) = \stiefel(n,p)/\mathrm{U}(p)$~\cite{absil2008}.

To optimize a generic function $f(A)$: $\grassmann(Dd,D) \to \mathbb{R}$ using any gradient-based optimization, we must be able to compute the gradient with respect to the parameters in $A$ and project it onto the tangent space of the Grassmann manifold.
The optimization of differentiable functions on Riemannian manifolds has been the object of extensive studies in mathematics and recently these techniques have been applied to TNs~\cite{hauru2021}.
For the self-containedness of this paper, we summarize the key ingredients of this optimization in Appendix~\ref{sec:tech}.

\begin{figure*}[thbp]
  \includegraphics{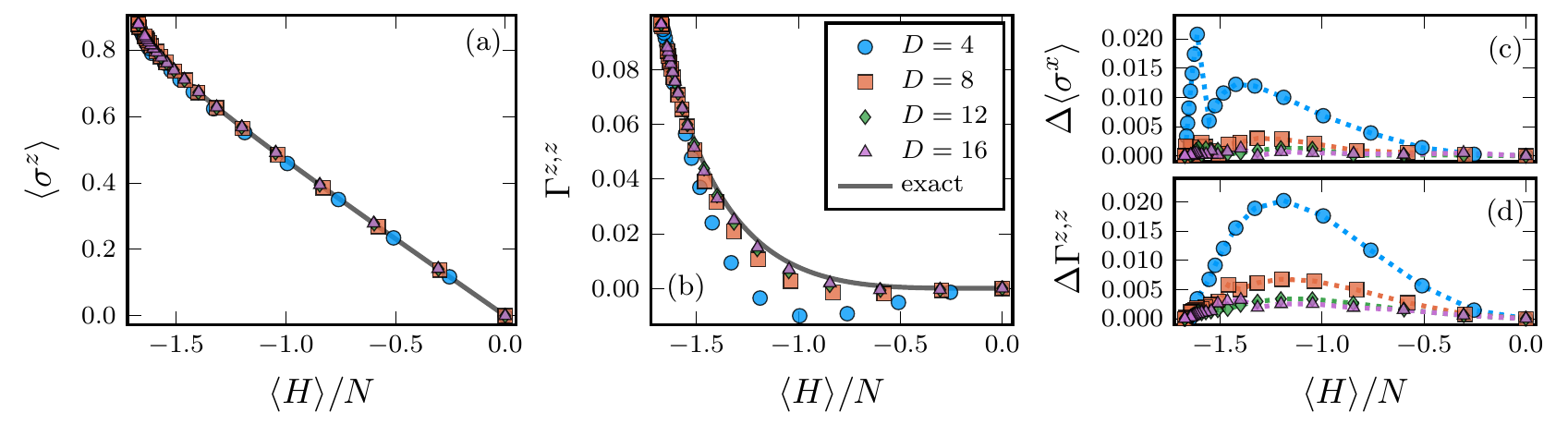}
  \caption{
    Magnetization (a) and next-neighbor correlation $\Gamma^{z,z}$ (b) versus the mean energy density for the Ising model with $h_x = 0$, $h_z = 3/2$, for different bond dimensions $D$.
    The dotted line corresponds to the exact results with the same mean energy density.
    In (c) and (d), the absolute errors to the exact solution are compared.
    No spontaneous symmetry breaking can occur at finite temperature in one-dimensional systems with local interactions---we therefore explicitly enforce the $\mathbb{Z}_2$ symmetry in the tensors.
  }
  \label{fig:integrable}
\end{figure*}
\begin{figure}[htpb]
  \includegraphics{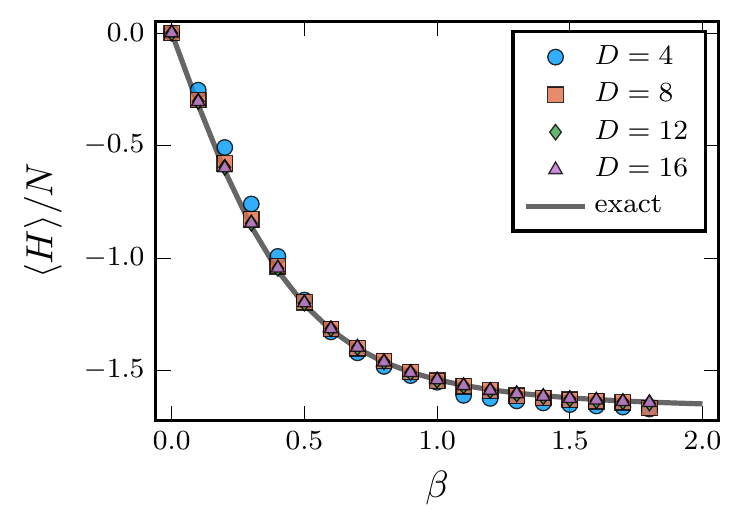}
  \caption{Average energy from Fig.~\ref{fig:integrable} choosing $\beta_R = \beta$.}
  \label{fig:integrable_beta} 
\end{figure}
\begin{figure*}[htpb]
  \includegraphics{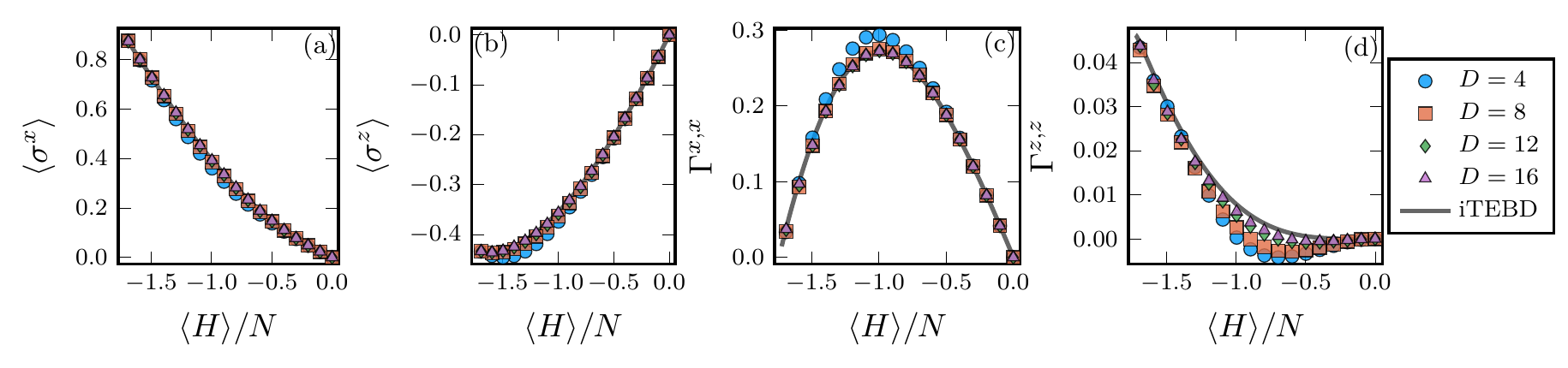}
  \caption{
    (a)--(d) Different observables as a function of the mean energy density for the nonintegrable case with $h_x = 0.5$, $h_z = -1.05$.
    The dotted line corresponds to the results given by the iTEBD algorithm.
  }
  \label{fig:nonintegrable}
\end{figure*}

For our application, the objective function is given by Eqs.~\eqref{eq:objectivefun}.
For the uniform MPS of Eq.~\eqref{eq:mps}, it reduces to
\begin{equation}
  f_R := \frac{F_R}{N} = \varepsilon + \frac{1}{\beta_R} \log \eta,
  \label{eq:objectivefun_mps}
\end{equation}
where $\varepsilon = \tr(H \rho)/N$ is the energy per site and $\eta = \left(\tr\rho^2\right)^{1/N}$ is the purity per site.
Both these terms are computable with standard TN routines in polynomial time, for uniform MPSs as well as finite MPSs.
The gradient of Eq.~\ref{eq:objectivefun_mps} with respect to $A$ is
\begin{equation}
  \frac{\partial f_R}{\partial A} = \frac{\partial \varepsilon}{\partial A} + \frac{1}{\beta_R \eta} \frac{\partial \eta}{\partial A}.
  \label{eq:gradient}
\end{equation}
As for Eq.~\eqref{eq:energy}, both these quantities $\partial \varepsilon/\partial A$ and $\partial \eta/\partial A$ are simple to obtain, as described in Appendix~\ref{sec:gradient}.
We thus use this gradient information to perform the optimization on the Riemannian manifold using the l-BFGS algorithm~\cite{liu1989,nocedal1999}.
An open-source implementation of the non-linear optimization in \textsc{Julia} is available online~\footnote{\url{https://github.com/giacomogiudice/RenyiOptimization.jl}}.

To conclude, we note that gradient methods cannot guarantee in any way convergence toward the global minimum, but only some local minimum.
While Eqs.~\eqref{eq:objectivefun} have a unique solution in the cone of the positive operators, the same cannot be said on a uniform MPS manifold of fixed bond dimension.

\subsubsection{Numerical experiments}
\label{sec:descent}
For our numerical experiments, we consider the Ising model:
\begin{equation}
  H = -\sum_k \left(\sigma^x_k \sigma^x_{k+1} + h_z \sigma^z_k + h_x \sigma^x_k \right).
  \label{eq:ising}
\end{equation}
When the parallel field vanishes ($h_x = 0$), the model is integrable, and local observables and correlations have a closed form~\cite{katsura1962,niemeijer1967}.
We use this model to perform the optimization of Eqs.~\eqref{eq:objectivefun} as described in Sec.~\ref{sec:optimization}.
The parameter $\beta_R$ is fixed to different values in the interval $\beta_R \in [0,2]$, and the uniform MPS is optimized until the gradient is sufficiently small~\footnote{The optimization halts after the norm of the gradient vector in tangent space is smaller than $10^{-6}$.}.

The results of the optimization are shown in Fig.~\ref{fig:integrable}, where we plot some local observables such as the magnetization $\braket{\sigma^z_i}$ and next-neighbor correlation $\Gamma^{a,b} = \braket{\sigma^a_{i} \sigma^b_{i+1}} - \braket{\sigma^a_{i}}\braket{\sigma^b_{i+1}}$ as a function of the mean energy density of the ensemble.
By increasing the bond dimension, we increase the number of the free parameters, and the numerical results converge toward the thermal ones.
Additionally, the comparison of the thermal observables by setting $\beta_R = \beta$ is shown in Fig.~\ref{fig:integrable_beta}.
Up to $\beta_R \lesssim 2$, we observe that there is a correspondence between the two ensembles at $\beta_R = \beta$.
For $\beta_R \gtrsim 2$, the optimization of Eq.~\eqref{eq:objectivefun_mps} converges to the ground space exactly, especially at small bond dimensions.
To study the physics of low temperatures, it is therefore more convenient to reexpress the optimization problem in Eqs.~\eqref{eq:objectivefun} by introducing a Lagrange multiplier:
\begin{equation}
  \rho^\ast = \arg \min_{\rho \succeq 0} \left\{\tr \rho^2 + \frac{\lambda^2}{2} \left( \tr(H \rho) - \bar{E} \right)^2\right\}.
\end{equation}
The gradient (see Appendix~\ref{sec:gradient}) can be modified accordingly, and the non-linear optimization can be performed in a similar way.
This objective function gets rid of the dependence on $\beta_R$, and one can directly choose an energy to target, since $\lim_{\lambda \to \infty}\tr(H\rho^\ast) = \bar{E}$.
However, if one wishes to explore the behavior of some observable with respect to $\bar{E}$, it is not necessary to perform the extrapolation with $\lambda \to \infty$, but a finite $\lambda$ is sufficient to obtain an energy in the vicinity of the desired value~\footnote{
  In our simulations, we set $\lambda = 10$.
  Since the purity per site $0.5 \leq \eta \leq 1$ is order 1, we expect deviations in energy density around $\order(1/\lambda^2)$.
}.

We also wish to remark that the method is completely general and does not depend on whether the system is integrable or not.
To complete our benchmarks, we present in Fig.~\ref{fig:nonintegrable} a comparison in the case where a parallel field is introduced, making the system non-integrable.
In this case, exact results are not known, but our results are compared to those of an MPS approximation to the Gibbs state purification obtained with a traditional imaginary time evolution method~\cite{vidal2003,verstraete2004}.
Since the model does not have a finite-temperature phase transition, the method will behave similarly for any value of the fields.
If one chooses $h_x=0$ and $h_z=1$, we expect that the required bond dimension increases when $\beta\to\infty$, as the critical ground state is approached~\cite{tagliacozzo2008,pollmann2009,pirvu2012,lauchli2013}.
In this regime, the cost function in Eq.~\eqref{eq:objectivefun} will be dominated by the energy term.
Hence the algorithm is reduced to an energy minimization, and we expect it to behave equivalently to other variational methods, such as the one proposed in Ref.~\cite{hauru2021}.

\subsection{Non-linear evolution}
\label{sec:evolution}
\begin{figure}[htb!]
  \includegraphics[width=0.8\columnwidth]{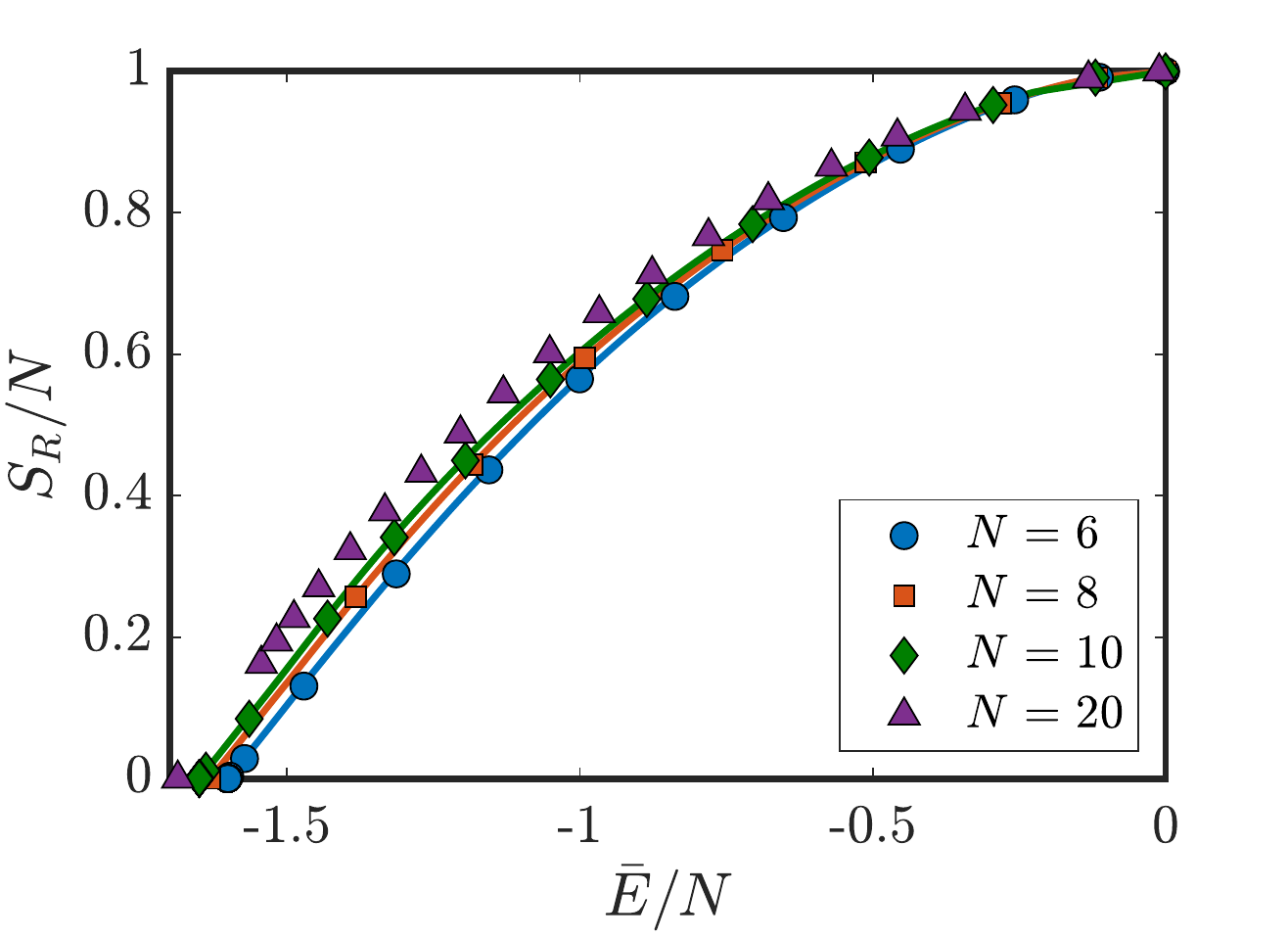}
  \caption{ 
    2-Rényi entropy of the maximal Rényi ensemble obtained with the analytic solution (solid lines) and nonlinear evolution (points).
    Results are for the Ising model (OBC) in Eq.~\eqref{eq:ising} with longitudinal and transverse fields, respectively, $h_x = 0.5$ and $h_z =-1.05$.
    We also show numerical results for $N=20$ (triangles) obtained using the non-linear evolution with MPS.
  }
  \label{fig:evolution}
\end{figure}
In Ref.~\cite{shi2020}, the authors introduced a non-linear evolution for approximating the thermal ensemble with Gaussian states.
Here we generalize this idea for the Rényi entropies, which gives rise to an evolution that is efficiently computable with TN techniques.
We consider a non-linear evolution of a density operator $\rho_\tau$ which depends on a real parameter $\tau$
\begin{equation}
  \dot{\rho}_\tau := \frac{\partial \rho_\tau}{\partial \tau} = - \frac{1}{2} \left\{\J_\tau - \braket{\J_\tau}, \rho_\tau\right\} .
  \label{eq:evolution}
\end{equation}
The operator $\J_\tau$ can be chosen such that the fixed point of this evolution gives rise to the MRE.
For example, the choice
\begin{equation}
  \J_\tau \rho_\tau = \beta_R H + \frac{2}{\tr \rho_\tau^2} \rho_\tau
  \label{eq:evolution_operator}
\end{equation}
gives rise to the same density operator as Eq.~\eqref{eq:objectivefun}.
The proof follows similarly from Ref.~\cite{shi2020}, and it is sufficient to show that the operator $\J_\tau$ in Eq.~\eqref{eq:evolution_operator} satisfies the following criteria:
\begin{subequations}
  \begin{align}
    &\tr{\rho}_\tau = 1, \; \forall \tau \in \mathbb{R} &\quad &\text{trace conservation,}
    \label{eq:evolution_criteria_trace}\\
    &\rho_\tau \succeq 0, \; \forall \tau \in \mathbb{R} &\quad &\text{positivity conservation,}
    \label{eq:evolution_criteria_positivity}\\
    &\partial f_R(\rho_\tau)/\partial \tau \leq 0 &\quad &\text{free-energy decrease.}
    \label{eq:evolution_criteria_decrease}
  \end{align}
  \label{eq:evolution_criteria}
\end{subequations}
Hence, choosing an appropriate density operator $\rho_0$ and integrating Eq.~\eqref{eq:evolution} over a sufficiently long interval, we obtain the solution to Eqs.~\eqref{eq:objectivefun}, since its value can only decrease with time.
There is no guarantee of reaching the global minimum---and indeed any eigenstate of $H$ does not evolve under Eq.~\eqref{eq:evolution}---but a random choice of the initial state should be sufficient in most cases.

We present some numerical experiments on small system sizes in Fig.~\ref{fig:evolution}, where the energy eigenbasis is available.
In all cases, the numerically integrated density operator converges to the ensemble in Eq.~\eqref{eq:mre}.
The evolution is discretized by expanding Eq.~\eqref{eq:evolution} to first order:
\begin{equation}
  \rho_{\tau + \delta \tau} \approx e^{-\frac{\delta \tau}{2} (\J_\tau - \braket{\J_\tau})} \rho_\tau e^{-\frac{\delta \tau}{2} (\J_\tau - \braket{\J_\tau})}.
  \label{eq:evolution_discretized}
\end{equation}
If the time step is chosen to be sufficiently small, then this evolution will converge to the desired fixed point.
This is witnessed by the fact that the Rényi entropy reaches the theoretical maximum for each mean energy, as shown in Fig.~\ref{fig:evolution}.
As a proof of concept, we also perform the integration using MPSs, in particular, using the TDVP scheme~\cite{haegeman2016,vanderstraeten2019} to update the state at each time step.
In practice, however, we observe that the time step required to obtain accurate results scales unfavourably with the system size, and we have yet to fully understand if the evolution becomes ill-conditioned for large system sizes.
Notwithstanding, it is possible that different integration schemes allow for large time steps without compromising the stability of the evolution.
We leave this as a venue for future work.

\section{Outlook}
\label{sec:conclusion}
In this paper, we have introduced an approach to compute thermal expectation values.
Instead of attempting to approximate the minimum of the free energy, we construct an ensemble that maximizes the 2-Rényi entropy for the same mean energy, and---in the thermodynamic limit---reproduces local observables of the corresponding Gibbs ensemble.

We have shown that this ensemble can be efficiently approximated using TNs and have presented variational algorithms to obtain such an approximation.
It is possible to work directly in the thermodynamic limit and use an MPS representation of the ensemble, which optimizes the objective function in Eqs.~\eqref{eq:objectivefun}.
Despite the simple form of this function, the optimization is non-linear and must be tackled with gradient-based methods.
The fundamental reason is that the positivity constraint in TNs is highly non-local, and one way of enforcing it is via a purification.
The convergence can be accelerated with techniques from manifold optimization, but a fundamental limitation is the high contraction cost.
Indeed, for a purification of bond dimension $D$, the time complexity involved in computing the purity (see Appendix~\ref{sec:gradient}) is $\order(D^5)$, which is significantly higher than the typical $\order(D^3)$ for other popular MPS algorithms, such as time evolution or ground-state search.
Coincidentally, the former is the same leading cost of the original formulation of DMRG with PBCs~\cite{verstraete2004a}.
Although the time complexity is higher, we observe that a moderate bond dimension captures well the ensemble and its local properties, both in integrable and nonintegrable models.

As an alternative to gradient-based optimization, we also propose an alternative method based on a non-linear evolution of the density operator.
Under this evolution, the objective function in Eqs.~\eqref{eq:objectivefun} is monotonically decreasing, and hence flows to the MRE.

Despite these limitations, we believe more efficient cost functions could be devised.
Additionally, the ideas outlined here could be applied to other wave-function ansätze.
For example, in recent works~\cite{yoshioka2019,hartmann2019,nagy2019,vicentini2019}, neural networks have been optimized with variational Monte Carlo to describe the steady state of dissipative dynamics.
Such techniques could be adapted to perform the optimization described in this paper.

\begin{acknowledgments}
We would like to thank E. Demler, F. Verstraete and T. Shi for useful discussions.
This work was partially funded by the Deutsche Forschungsgemeinschaft (DFG, German Research Foundation) under Germany's Excellence Strategy---EXC-2111---390814868 and by the European Union through the ERC grant QUENOCOBA, ERC-2016-ADG (Grant No. 742102).
\end{acknowledgments}

\bibliography{library}

\appendix
\section{Calculation of the variance for a Gaussian density of states}
\label{sec:variance}

\begin{figure*}[htbp]
  \includegraphics{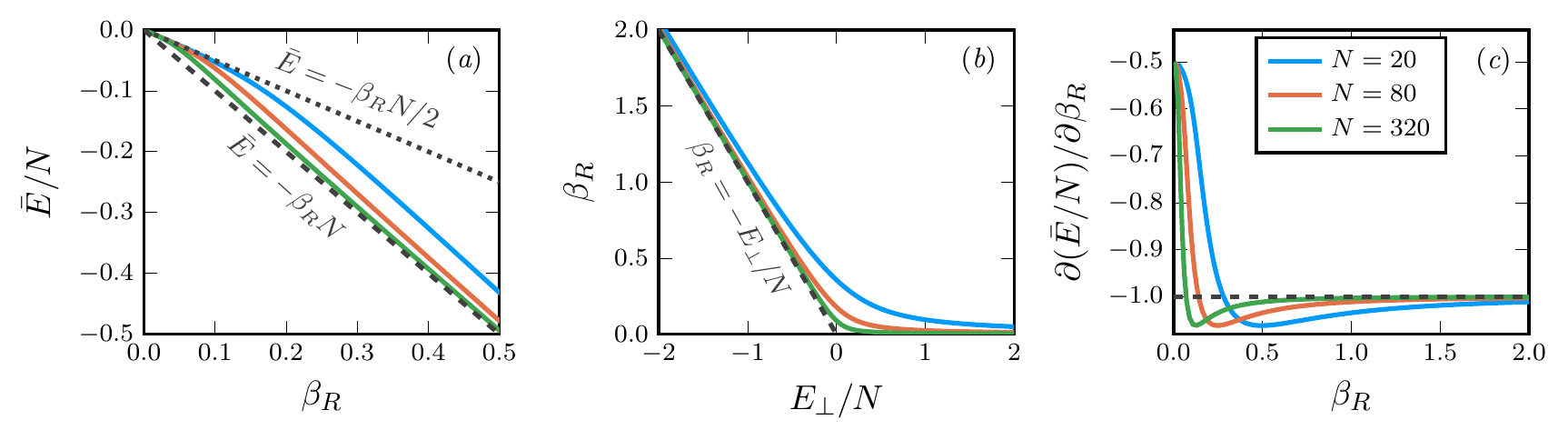}
  \caption{
    (a) The energy density of the MRE as a function of the parameter $\beta_R$, for increasing system sizes.
    For clarity, in all plots we set $\sigma^2 = 1$.
    In the high-temperature limit ($\bar{E} \to 0$), the behavior $\bar{E} = \beta N/2$ (dotted line) indicates that $\beta_R \approx 2\beta$.
    However, as the system size is increased, at non-zero energy density the equation of state approaches $\bar{E} = \beta N$, which is analogous to the Gibbs ensemble.
    Hence, in the limit of $N \to \infty$ we can identify $\beta_R \approx \beta$.
    This can be understood in (b), for which $\beta_R \approx \max(E_\perp/N, 0) + \order(1/\sqrt{N})$.
    (c) Taking the derivative $\partial (\bar{E}/N)/\partial \beta_R$, we identify a fast-varying regime, corresponding to $E_\perp < 0$, and a second regime at $E_\perp > 0$ where the derivative is close to $-1$.
  }
  \label{fig:gaussian}
\end{figure*}

In this appendix, we compute the mean energy $\braket{H}$ and variance $\var{H} = \braket{H^2} - \braket{H}^2$ for the Gibbs and 2-Rényi ensembles, assuming the density of states is a Gaussian of the form
\begin{equation}
  \dos(E) = \exp \left(-\frac{(E - E_\mathrm{mid})^2}{2\sigma^2 N}\right),
  \label{eq:dos_gaussian}
\end{equation}
where $N$ is the system size and $\sigma$ is a constant independent of $N$.
Additionally, without loss of generality, let us assume it is centered at $E_\mathrm{mid} = 0$.

For local Hamiltonians as Eq.~\eqref{eq:hamiltonian}, it was shown that the density of states weakly converges to a Gaussian in the thermodynamic limit, as a consequence of Lyapunov's central limit theorem~\cite{hartmann2004,hartmann2005,keating2015}.
As an alternative proof, one can take a an ancillary copy of the system, and consider the state $\ket{\Xi}$, which is the tensor product of maximally entangled pairs between system and ancilla:
\begin{equation}
  \ket{\Xi} = \bigotimes_n \frac{1}{\sqrt{|\hilb_n|}} \sum_{i=1}^{|\hilb_n|} \ket{i}_\text{sys} \ket{i}_\text{anc}.
\end{equation}
In the doubled system, the state $\ket{\Xi}$ is a product state, and one can apply directly the Theorem in Ref.~\cite{hartmann2004} to obtain the desired result.

However, the rate of convergence to the central limit theorem is larger than $\order(1/\sqrt{N})$, and one should take into account the finite-size corrections when computing expectation values.
Hence, we can think of Eq.~\eqref{eq:dos_gaussian} as a toy model of actual local Hamiltonians, and derive results under this assumption.

For the Gibbs ensemble, we have that the partition function is
\begin{equation}
  \Z_G = \int_{-\infty}^{+\infty} \!\! e^{-\beta E} D(E)\,\dd E.
\end{equation}
This leads to
\begin{subequations}
  \begin{align}
    \braket{H}_G &= -\frac{\partial}{\partial \beta}\log \Z_G = -\beta \sigma^2 N,\\
    \var{H}_G &= \frac{\partial^2}{\partial \beta^2}\log \Z_G  = \sigma^2 N.
  \end{align}
\end{subequations}
Naturally, these results hold only in the region around the peak of the Gaussian, and break down when one tries to take the limit of $\beta \to \infty$.

For the 2-Rényi distribution, we cannot use the trick of deriving the partition function with respect to $\beta$, since we cannot interpret it as a generating function.
We can however express everything in terms of the truncated moments:
\begin{equation}
  \Phi_m := \int_{-\infty}^{E_\perp} \!\! E^m D(E)\,\dd E.
\end{equation}
The upper integration limit is related to the mean energy and $\beta_R$ as $E_\perp = \bar{E} + \frac{2}{\beta_R}$.
These moments enjoy a recurrence relation of the form $\Phi_{m+2} = \sigma^2 \partial \Phi_{m}/ \partial \sigma$.
Additionally, $\Phi_1$ is analytical because the integrand is the derivative of a Gaussian.
This allows us to establish the identities
\begin{subequations}
  \begin{align}
    \Phi_2 &= \sigma^2 N \Phi_0 + E_\perp \Phi_1, \\
    \Phi_3 &=  \left(2 \sigma^2 N + E_\perp^2\right)\Phi_1.
  \end{align}
\end{subequations}
By dividing the partition function by $\beta/2$, we can then compute the mean energy for this ensemble as
\begin{equation}
  \braket{H}_R = \frac{E_\perp \Phi_1 - \Phi_2}{E_\perp \Phi_0 - \Phi_1}
  = - \sigma^2 N \frac{\Phi_0}{E_\perp \Phi_0 - \Phi_1}.
\end{equation}
Equating this result to $\bar{E}$ allows us to express $\Phi_0$ in terms of $\Phi_1$:
\begin{equation}
  \Phi_0 = \frac{\bar{E}}{\bar{E} E_\perp + \sigma^2 N} \Phi_1.
\end{equation}
Using this last relation, we can write the variance as
\begin{equation}
  \var{H}_R = \bar{E} E_\perp + 2 \sigma^2 N - \bar{E}^2 = 2 \sigma^2 N + \frac{2 \bar{E}}{\beta_R}.
\end{equation}

For the Gibbs ensemble, notice that $\beta$ and $\bar{E}$ are collinear, $\bar{E} = - \beta \sigma^2 N$.
At infinite temperature ($\bar{E}=0$), we have trivially $\beta = \beta_R = 0$.
Expanding $\rho_G$ around $\beta = 0$, we obtain $\rho_G \approx 1 - \beta H$.
Comparing this with the form of the MRE, we can easily conclude that in this limit $\beta_R \approx 2 \beta$.
However, one should take into account the thermodynamic limit.
Indeed, as shown in Fig.~\ref{fig:gaussian}, at non-zero $\bar{E}$, increasing the system size leads to an equation of state which asymptotically approaches $\bar{E} = -\beta_R \sigma^2 N$.
This is due to the fact that the cutoff $E_\perp$ becomes proportional to $\beta_R$.
Indeed, at $E_\perp = 0$, one has that $\beta_R(E_\perp = 0) = \order(1/\sqrt{N})$.
The point $E_\perp = 0$ also corresponds to a stationary point of $\partial \bar{E}/\partial \beta_R$.
Taking derivatives, one obtains a relation between $\bar{E}$ and $\beta$ only
\begin{equation}
  \frac{\partial \bar{E}}{\partial \beta_R}
  = \frac{\partial \bar{E}}{\partial E_\perp} \frac{\partial E_\perp}{\beta_R}
  = \frac{\sigma^2 N}{\beta_R^2}\frac{1 + 2\frac{\bar{E}}{\beta_R \sigma ^2 N}}{1 +\frac{\bar{E}}{\beta_R \sigma^2 N}}.
\end{equation}
As the system size is increased, the derivative converges toward a constant, as shown in Fig.~\ref{fig:gaussian}.
This allows us to conclude that, for a Gaussian density of states and $\beta_R \gg 1/\sqrt{N}$, we have $\beta_R \approx \beta$.

\section{Technical details on Grassmann manifolds}
\label{sec:tech}
The gradient of a function on a Riemannian manifold belongs to the tangent space of the manifold itself.
A generic tangent vector to a uniform MPS is a linear combination of the partial derivative with respect to the single tensor.
This can be seen as a vector embedded in Hilbert space, composed of an (infinite) sum of MPS vectors
\begin{multline}
  \ket{\Delta(B)} = B \frac{\partial}{\partial A} \ket{\Psi(A)} \\
  = \sum_n \includetikz{tangent}
\end{multline}
where the sum runs over all physical sites.

A tangent vector parametrized by a tensor $B$ has an inherent gauge freedom to it.
The explicit transformation that leaves the vector invariant is $B^s \mapsto B^s + X A^s - A^s X$, for any $D \times D$ matrix $X$.
Indeed, the set of derivatives $\ket{\partial_\mu\Psi(A)}$ form an overcomplete basis.
Hence, by introducing the orthogonal complement of A~\cite{hauru2021}, such that
\begin{equation}
  \includetikz{complement}
\end{equation}
we can parametrize the tangent vectors as
\begin{equation}
  \includetikz{parametrization},
  \label{eq:grassmann_parametrization}
\end{equation}
where $Z$ is a $D(d-1) \times D$ matrix.
This parametrization arises quite naturally if one considers the tangent vectors to be embedded in the original Hilbert space.
In this case, the choice of Eq.~\eqref{eq:grassmann_parametrization}
corresponds to imposing orthogonality of the tangent space, $\braket{\Psi(A)|\Delta(B)} = 0$.
We remark that Eq.~\eqref{eq:grassmann_parametrization} is exactly the parametrization of the tangent space for Grassmann manifolds, as derived traditionally~\cite{absil2008}.

Hence we can consider the problem Eq.~\eqref{eq:objectivefun} as an optimization of a tensor $A$ over the Grassmann manifold.
One particularity arises from the choice of metric in the tangent space.
In Riemannian manifold optimization, one usually chooses the Euclidean metric~\cite{absil2008}:
\begin{multline}
  \braket{\Delta_1(B_1) | \Delta_2(B_2)}_{\text{Eucl}} = \real\tr(B_1^\dag B_2) \\
  = \real\!\includetikz{metric-euclidean}.
  \label{eq:metric_euclidean}
\end{multline}
However, this is not the most natural choice in this setting, since the underlying physical Hilbert space prescribes the metric
\begin{multline}
  \braket{\Delta_1(B_1) | \Delta_2(B_2)}_{\text{Hilb}} \propto \real\tr(B_1^\dag B_2 \varrho) \\
   =  \real\includetikz{metric-hilbert}.
  \label{eq:metric_hilbert}
\end{multline}
Note that, as opposed to Eq.~\eqref{eq:metric_euclidean}, this metric depends on the current point $\ket{\Psi(A)}$ of the manifold.
In practice, we notice that the choice of the metric is not very important for the optimization, and the Euclidean metric poses the advantage of not having to invert a potentially ill-conditioned fixed point $\varrho$ when projecting onto the manifold.
Additionally, the use of the Euclidean metric is not necessarily deleterious since, compared to Eq.~\eqref{eq:metric_hilbert}, it will magnify the importance of small Schmidt values of the state $\ket{\Psi(A)}$.
In Ref.~\cite{hauru2021}, the authors have proposed a non-linear preconditioner that acts as a compromise between these two metrics.
Regardless of our choice, the metric allows us to project arbitrary Hilbert space vectors onto the tangent space.
The projection operator for the Euclidean metric in Eq.~\eqref{eq:metric_euclidean} reads~\cite{edelman1998}
\begin{equation}
  \proj_{A}(Y) = \includetikz{project}.
  \label{eq:grassmann_projector}
\end{equation}

For the optimization of a generic function $f(A)$: $\grassmann(Dd, D) \to \mathbb{R}$, we can compute the gradient $\nabla f(A)$ without taking into account the isometricity condition and then projecting onto the tangent space~\cite{absil2008}.
This projected gradient $\grad f(A) := \nabla f(A)$, can be considered the direction of steepest ascent on the manifold, while its magnitude can be used as a convergence criterion.

The last ingredient necessary for a gradient descent algorithm is defining a \emph{retraction}.
Loosely speaking, we need to define a curve on the manifold such that we can move in a direction specified by a tangent vector by a step size $\alpha$.
Hence a retraction $\retr_{A}(\alpha \Delta)$ can be any smooth curve such that the it (i) starts at $A$, $\retr_{A}(0) = A$ and (ii) is consistent with $\dd \retr(\alpha \Delta) /\dd \alpha = \Delta$.
Different choices of retraction exist, but the most natural choice is a retraction that follows the manifold geodesics, i.e.,\ the shortest path that connects two points on the manifold.
Remarkably, the geodesic retraction on a Grassmannian manifold relative to the Euclidean metric is known and is relatively efficient to compute~\cite{edelman1998}: Given some point $A$, the retraction of some tangent vector $\Delta = A^\perp Z$ [see Eq.~\eqref{eq:grassmann_parametrization}] is
\begin{equation}
  \label{eq:grassmann_retraction}
  \retr_{A}(\alpha \Delta) = e^{\alpha Q} A,
\end{equation}
where
\begin{equation}
  Q =
  \begin{pmatrix}
    A & A^\perp
  \end{pmatrix}
  \begin{pmatrix}
    0 & -Z^\dag \\
    Z & 0
  \end{pmatrix}
  \begin{pmatrix}
    A^\dag \\
    {A^\perp}^\dag
  \end{pmatrix} .
\end{equation}

This constitutes the bare minimum to define a gradient descent algorithm on the Grassmann manifold.
In practice, the convergence of gradient descent can be very slow, and, in Euclidean space, several methods that just use first-order information.
For example, conjugate gradient adjusts the gradient with the previous search direction, and quasi-Newton methods---notably l-BFGS~\cite{liu1989,nocedal1999}---uses the previous iterations to create a low-rank approximation of the inverse Hessian.
To adapt these methods to optimization on manifolds, it is sufficient to define a \emph{vector transport}, a way of transporting a tangent vector at a previous point of the manifold to the current one.
In other words, for a retraction $A' = \retr_{A}(\Delta)$, a vector transport $\transp_{\Delta}(\Omega)$ maps a tangent vector $\Omega$ at $A$ to a tangent vector at $A'$.
A typical way of defining transport is via differentiated retraction, i.e.,
\begin{equation}
  \transp_{\Delta}(\Omega) = \left.\frac{\dd}{\dd \alpha} \retr_{A}(\Delta + \alpha \Omega)\right|_{\alpha = 0}.
\end{equation}
Using Eq.~\eqref{eq:grassmann_retraction}, we obtain
\begin{equation}
  \label{eq:grassmann_transport}
  \transp_{\alpha \Delta}(\Omega) = e^{\alpha Q} \Omega.
\end{equation}

\section{Calculation of the gradient}
\label{sec:gradient}
For simplicity, we assume that the Hamiltonian is two-local and $h_n = h\ \forall N$, but the algorithm can be readily generalized to a non-trivial unit cell and any Hamiltonian which has an MPO form~\cite{zauner-stauber2018}.
Once the leading fixed point $\varrho$ is computed, the energy density reduces to the following network:
\begin{equation}
  \varepsilon = \includetikz{energy}.
  \label{eq:energy}
\end{equation}
To compute the derivative $\partial \varepsilon/\partial \bar{A}$ of Eq.~\eqref{eq:energy} it is useful to define the left and right environments corresponding to the geometric sum of the terms in the Hamiltonian over each half-infinite chain:
\begin{widetext}
  \begin{equation}
    \includetikz{energy-left} = \includetikz{energy-open-left},
    \qquad
    \includetikz{energy-right} = \includetikz{energy-open-right},
    \qquad
    \mathbb{E} = \includetikz{transfer}.
  \end{equation}
  The notation $(1 - \mathbb{E})^{\rm P} = (1 - \mathbb{E} + \pket{\varrho}\pbra{\id})^{-1}$ is used to denote the geometric where the divergent part---corresponding to the leading eigenpair---is subtracted~\cite{zauner-stauber2018}.
  Hence the gradient $\partial \varepsilon/\partial \bar{A}$, without accounting for the constraint is
  \begin{equation}
  	\frac{\partial \varepsilon}{\partial \bar{A}} = \includetikz{fixedpoint-gradient-left} + \includetikz{energy-gradient-left} + \includetikz{energy-gradient-right} + \includetikz{fixedpoint-gradient-right}.
  	\label{eq:gradient-energy}
  \end{equation}
  When computing the purity, we can retain the leading eigenvalues and eigenvectors of the transfer element of $\rho^2$:
  \begin{equation}
    \includetikz{doublefixedpoint-left} = \eta \includetikz{doublefixedpoint-transfer-left},
    \qquad
    \includetikz{doublefixedpoint-right} = \eta \includetikz{doublefixedpoint-transfer-right},
    \qquad
    \includetikz{doublefixedpoint-norm} = 1.
  \end{equation}
  This is by far the costliest computational step relative to the bond dimension $D$, since it scales as $\order(D^5)$, as opposed to the other steps which are all $\order(D^3)$.
  The gradient $\partial \eta/\partial \bar{A}$ becomes
  \begin{equation}
  	\frac{\partial \eta}{\partial \bar{A}} = \includetikz{doublegradient-upper} + \includetikz{doublegradient-lower}.
  	\label{eq:gradient-purity}
  \end{equation}
  We can then put together Eq.~\eqref{eq:gradient-energy} and Eq.~\eqref{eq:gradient-purity} to obtain the gradient $\partial f_R/\partial \bar{A}$ in Eq.~\eqref{eq:gradient}.
  To compute the gradient on the Grassmann manifold, we must then project the unconstrained gradient using Eq.~\eqref{eq:grassmann_projector}.
\end{widetext}
\end{document}